\documentstyle[aps,prd,preprint]{revtex}
\def\gtorder{\mathrel{\raise.3ex\hbox{$>$}\mkern-14mu
             \lower0.6ex\hbox{$\sim$}}}
\def\ltorder{\mathrel{\raise.3ex\hbox{$<$}\mkern-14mu
             \lower0.6ex\hbox{$\sim$}}}

\begin{document}
\draft
%
%\hfill CITA-96-19
\title{Evolution of the Cosmic Gas and the
 Relic Supernova Neutrino Background}
\author{R.A.\ Malaney}
\address{Canadian Institute for Theoretical Astrophysics, University of Toronto,\\
Toronto ON Canada M5S 1A7}

\date{\today}

\maketitle
\begin{abstract}

Using the redshift evolution of the cosmic gas, as inferred from
QSO absorption line studies, we  calculate the past supernova rate and
the relic supernova neutrino background. Using this new technique we
find the predicted relic neutrino flux at low energies to be at least
an order of magnitude below previous estimates. We argue that
the evolution of the cosmic gas is consistent with a large decrease
in the number of early-type galaxies at redshifts $\sim 3$,
and that this evolution is the source of the reduction in the
predicted neutrino flux. Additional observational constraints from
recent redshift surveys lead us to propose a modified model for the 
evolution of the cosmic gas in which significant infall at low redshift
occurs. We discuss the possible relevance of our calculations to 
the X-ray emitting metal-enriched gas observed in the intergalactic
medium.

\end{abstract}

\pacs{
}

\section{Introduction}

All available evidence seems to support the idea that the
damped Ly$\alpha$ systems are the progenitors of  current galaxies
\cite{LWT,WLF}. Also,
measurements of the H~I column densities in these systems allow
 the redshift-evolution of the neutral gas in the universe to be
 determined \cite{LWT,WLF,stor}. 
If, as  argued by  \cite{LWT}, the evolution
of this gas is solely due to the transformation of gas into stars and vice versa
(i.e. a closed box system), then a rather direct inference of the star formation
rate  (SFR) and chemical evolution at high redshift can be made \cite{LWT}.
Other studies which relax the closed box assumption, and which
probe the possible influence of dust on the  predicted chemical evolution,
 have also been carried out \cite{dust}.

A principle aim of this work is to utilize  the  inferred evolution of
 the cosmic gas in order to ascertain the rate of core-collapse supernova as
a function of redshift, and to use this information  to determine the relic
supernova neutrino background.   Due to its potential to provide  insight
into the formation history of galaxies,
the relic supernova neutrino background
has been the focus of several previous studies \cite{russ,krau,woos,tota,tot}. 
Interest in this subject 
has been heightened
 due to the recent completion
of the Superkamiokande neutrino detector \cite{SK}, which will attempt for
the first time a direct measurement of the MeV neutrino background.

  By comparing our new  predictions  of the supernova rate with
 previous results based on present day galaxy counts,
 we shall attempt to provide
some insight into the evolution of galaxies. Also, from comparisons 
with the luminosity densities obtained from the CFRS and Hubble Deep Field
surveys, we shall explore the possibility that the evolution of the cosmic gas
at low redshift requires modifications from a closed-box description.

In this report we will be extending 
the numerical calculations of the  cosmological SFR
and chemical evolution   presented in a previous
report \cite{me}.
 Unlike the  earlier work of  \cite{LWT,dust}, the calculations
reported here do not assume analytical solutions
based on the instantaneous recycling approximation (IRA).
In the work of \cite{me} it was shown that the error in the
predicted cosmic SFR introduced by the use of the IRA was   
$\ltorder 20\% $ (for
most initial mass functions).
However, 
in order to identify the
cosmological initial mass function -- knowledge of which is a prerequisite
to determining the universal supernova rate -- 
 a detailed comparison of the predicted and
  observed abundances is required. 
For various reasons, this program is best carried out using
 numerical calculations of the type reported here.
For further details on  the calculations,
beyond that given below,
 the reader is pointed to \cite{me}.

\section{THE PAST SUPERNOVA RATE AND RELIC NEUTRINOS}

The cosmological gas density $\Omega_{g}(z)$ 
 is defined as the comoving
gas mass density  at a redshift  $z$ in units of the present
 critical density $\rho_c$.  Fig.~1 shows the inferred values of
$h\Omega_{g}(z)$  for $q_{\rm 0}=0.5$ as tabulated by  Storrie-Lombardi
{\it et al.}
\cite{stor} (the Hubble constant is given by 
$H_{\rm 0}=h100$ km s$^{-1}$  Mpc$^{-1}$). The curves represent
 parameterizations of the data
using the functional formalism outlined by Pei and Fall \cite{dust}.
The solid curve represents a parameterization to the most recent
data (solid circles) \cite{stor}, whereas the dotted curve
represents a parameterization to the 
data of Lanzetta  {\it et al.} \cite{LWT}. In the calculations to follow
we will adopt these two curves as  a coarse indicator
of the uncertainty in the gas evolution. Their use also allows us to explore how
 variations in  $\Omega_g(z)$ affect our calculations.
Uncertainty  in  $\Omega_g(z)$ is  even more
warranted if dust  complicates the issue.
The  dashed curve  of Fig.1 indicates
 a possible correction to the dotted curve if  observational bias
introduced by dust  effects are important, as suggested by \cite{dust}.
Although the dust-corrected  gas  evolution 
is difficult to determine accurately, the dashed curve  can be
taken as an  example of its potential effects.
For the time being we shall ignore
 the incomplete dot-dashed curve of Fig.1.
This does not represent any fit to the data,
and is used later for illustrating the effects of
very  high supernova rates.

 In the closed box model the evolution of $\Omega_{g}$
 with redshift $z$ is described by
\begin{equation}
{d\Omega_{g}\over dz}=   \biggl \{\ { \int^{m_{up}}_{m(z)} (m-m_r)
 \Psi(z^f_m) \Phi(m) dm
-\Psi(z) } \biggr \} {dt\over dz}  \ \ \ .
\end{equation}
Here $z^f_m$ is the formation  redshift at which a star of mass $m$ is
returning gas back to the interstellar medium at the current redshift
 $z$, $\Psi$ is
the SFR, $\Phi$ is the initial mass function (IMF), and $m_r$ is the remnant
stellar mass.
The lower limit of integration $m(z)$ is the minimum  stellar mass which
can be returning gas to the system at redshift $z$. 
We assume a power law for the IMF viz. $\Phi(m) \propto m^{-(1+x)}$, 
 normalized through the relation
\begin{equation}
\int^{m_{up}}_{m_{low}} m \Phi(m)dm=1 \ \ \ ,
\end{equation} 
where $m_{up}=40 {\rm M}_\odot$ and
$m_{low}=0.08 {\rm M}_\odot$ (experiments where we adjust the upper
limit to $100 {\rm M}_\odot$, suggest  factor $\sim 2$
 modifications to most of our
predicted observables).
Assuming  the cosmological  constant is zero
 (as we will in all our calculations), the  cosmic
 time $t$ is related to $z$ through
\begin{equation}
{dt\over dz}=-\biggl \{H_{\rm 0}  (1+z)^2 \sqrt{(1+\Omega_{\rm 0} z)} \biggr \}^{-1}
\ \ \ ,
\end{equation}
where $\Omega_{\rm 0}=(\rho/\rho_c)_{\rm 0}$, $\rho$ being the energy density.
Unless explicitly stated otherwise, we assume $\Omega_{\rm 0}=1$.

The rate of core-collapse
supernova,  $R_{\rm SN}$ 
 can be determined through
\begin{equation}
R_{\rm SN}=\rho_c 
  \int^{m_{up}}_{m16(z)}  \Psi(z^f_m) {\Phi(m)} dm
+(1-k) \rho_c\int^{16}_{m11(z)}  \Psi(z^f_m) {\Phi(m)} dm \ \ \ ,
\end{equation}
where the lower limits of integration  $m16$ and $m11$ 
have  the same meaning as before, except they
 have lower limits of $16$ and $11 {\rm M}_\odot$, respectively.
The quantity $k$ is a small number  related to
the relative number of Type~I supernovae (we will take $k= 0.005$,
see \cite{me}). 
For the purposes at hand its adopted value is unimportant 
(eg. adopting
 $k=0$  introduces a negligible effect on the neutrino flux).
 With $\rho_c$ in units of $m$ per 
 volume, and time in  say Gyr,
Eq.~(4) gives the  comoving supernova rate in  dimensions  of number 
per Gyr per volume.

Let us focus on the 
electron anti-neutrinos $\bar\nu_e$, which are the
neutrinos Superkamiokande will attempt to detect.
Given the temperature and total energy of the emitted $\bar\nu_e$'s, and
assuming a Fermi-Dirac  energy distribution with zero chemical potential, 
the 
number of $\bar\nu_e$'s of energy $q$ per unit $q$, ${dN[q]\over dq}$,
emitted by a supernova can be determined. Here, we adopt the 
temperature and energy values tabulated  as a function
of progenitor mass by Woosley  {\it et al.} \cite{woos} to arrive at
a weighted average for ${dN[q]\over dq}$, which we call  ${d\bar N[q]\over dq}$.
Clearly this weighted average will
depend on the adopted IMF, and in the relations to follow one should take
${d\bar N[q]\over dq}$ to be averaged over the given IMF.

One can  show that the differential number flux
 of
 relic supernova
 $\bar\nu_e$'s    at the present epoch 
can be given in terms of the above quantities as
\begin{equation}
{dF_\nu\over dq} = c\int^0_z R_{\rm SN}(z) {d\bar N[q']\over dq} (1+z) {dt\over dz} dz
\ \ \ \ ,
\end{equation}
where $q'=(1+z)q$ and $c$ is the speed of light.
Explicitly in terms of the SFR  and IMF this becomes 
\begin{equation}
{dF_\nu\over dq} = {3cH_{\rm 0}\over 8\pi G} \int^z_0 \biggl \{ 
  \int^{m_{up}}_{m11(z)}  \Psi(z^f_m) {\Phi(m)} dm \biggr \}
{d\bar N[q']\over dq}
\biggl \{(1+z)\sqrt{(1+\Omega_{\rm 0} z)}\biggr \}^{-1}
 dz
\ \ \ \ \ ,
\end{equation}
where we have substituted in for $\rho_c$ in order to explicitly
show the linear dependence of the flux on $H_{\rm 0}$
($G$ is the gravitational constant).
Although included in the calculations,
for clarity we have dropped in Eq.~(6) the term proportional to $(1-k)$.

In order to proceed,
 we must calculate the supernova rate as a function of redshift,
which requires that we specify the  cosmic IMF to be adopted.
For a given evolution of the cosmic gas, the IMF can be chosen
so as to correctly reproduce the observed cosmic metallicity
at a particular redshift. (Many studies of the metal abundances in damped
Ly$\alpha$ systems have now  appeared, the most recent of which
can be found in \cite{keck}.)
 Using the evolutionary curve corresponding
to the data of  Lanzetta {\it et al.} \cite{LWT}
 (dotted curve of Fig.~1) and normalizing
to solar metallicity at redshift zero,
 the cosmic IMF is best described by a slope parameter $x=1.7$ \cite{me}.
Henceforth, for calculations in which we utilize the dotted (or dashed) curve of
Fig.1, we will assume $x=1.7$.
However,
modeling the cosmic gas evolution  from  the recent data of 
Storrie-Lombardi {\it et al.} \cite{stor} (solid curve of
Fig.~1), allows the IMF slope parameter to be reduced slightly.
This is due to the reduced level of star formation predicted by this
new data, which in turn requires a flatter IMF for
a given level of metal production.  With the new $\Omega_g(z)$
evolution
a Salpeter IMF, $x=1.35$, in fact
 adequately reproduces the abundance data.
Henceforth, we shall refer to a closed-box model
  using the $\Omega_g(z)$
evolution signified by the solid curve of Fig.~1, and an IMF slope $x=1.35$,
 as our {\it standard}  chemical evolution model.

 That our standard model accommodates the abundance data is seen more explicitly
 from 
Table~1, which lists the abundance mass fractions in the cosmic gas at various
redshifts (here $Z$ is the total metallicity).
As will become clearer later, there are  doubts as to whether
the closed box approximation at $z\ltorder 1.5$ is valid. 
For this reason it is best to normalize the IMF at higher redshifts,  $z=2.5$ say.
This redshift also has the advantage of being well sampled.
Since its depletion onto grains is anticipated to be relatively small,
Zn is  widely
used as a tracer of the gas metallicity. From Table~1 we find that 
the Zn abundance of our standard chemical
evolution model
 is predicted to be approximately
0.03 solar at $z=2.5$, and is
consistent
with the mean Zn abundance  given by   {\cite{keck,pett2}
 (see also later discussion of Fig.6).
Thus, we are comfortable that   this model is in
reasonable agreement with the present abundance data of damped Ly$\alpha$
systems.
Whether or not the
 other element abundances given in Table 1 represent the observed mass
fractions in the cosmic gas,
 depends on the relative importance of
grain depletion effects. Comparing  directly with the abundances 
measured by \cite{keck}, the predictions of Table~1
 suggest a  grain depletion  factor of order three
for Fe at $z=2.5$.

Using the evolutionary curves and the IMF's just described, 
Eqs.(1)-(4) lead to the supernova rates shown in Fig.2.
The curve markings  correspond to those shown in Fig.~1, except that
we  also show the rate predicted by a calculation in which our standard
 chemical evolution model is modified by the inclusion of a late
infall term (long-dashed curve). This  infall model will be explained in detail later.
For convenience we have plotted the rate
as the number of supernovae per year in a galaxy of mass
 $10^{11}{\rm M}_\odot$. 
Doing this requires that the rates shown in Fig.2
be
multiplied by $0.01/\Omega_{gal}$, where $\Omega_{gal}$ is the mass density
(of ordinary matter)
in galaxies relative to the critical density.
Note that  in defining the rates this way, the  explicit
dependencies on $H_{\rm 0}$  cancel
out. We also note that except for the late-infall model,  
the predicted supernova rates at low redshift are quite low. For example, 
our standard chemical evolution model leads to a predicted
supernova rate at $z=0$ of approximately one per $350$ years per
$10^{11}{\rm M}_\odot$ galaxy (for $\Omega_{gal}=0.007$).
 Although quite uncertain, the  observationally inferred  rate
for a galaxy typical of the Milky Way
 is thought to be
in the range of 1 per 30-100 years \cite{berg,tam}.
The theoretical supernova rate for a typical $10^{11}{\rm M}_\odot$  galaxy
 after averaging over all morphological galaxy
types  is  also in  the 1 per 100 years range, 10~Gyr after 
galaxy formation commenced \cite{tot}.

One could take the predicted low supernova rate at zero redshift to indicate
a breakdown  in the extrapolation of the last data point
toward zero redshift. This could be caused by any one of several effects
occurring within the last data bin 
such as  coarse  averaging,
 galaxy evolution, or a breakdown of the closed box approximation (see later).
 Whatever the reason, we view   the predicted low  supernova
 rates   at zero redshift with suspicion.
To compensate for this, 
we adopt  a procedure for some of our calculations in which
we set the   supernova rate to some constant rate, $r_c$, 
for $z<z_c$, where $z_c$ is the redshift at which our 
 predicted rate falls below $r_c$.
This constant rate is set to $r_c= a(0.01/\Omega_{gal})$ per
 $100$ years  per
$10^{11}{\rm M}_\odot$ galaxy  ($0.5 \le a \le 2$).

Fig.3 displays the predicted neutrino spectrum for a variety of models.
The solid curve represents the spectrum arising from our
standard chemical evolution model (no low redshift $r_c$
correction). Three of the curves correspond
to  a modification (in the manner outlined above) to the  supernova rate predicted by
our standard chemical evolution model. For the short-dashed curve
we have adopted a value of $r_c$ with $a=1$.
Adjusting the constant rate,
the lower (higher) dotted  curve  
corresponds to a decrease (increase) of
$r_c$ at low redshift by a factor two.
 The long-dashed curve is the prediction from our late infall model.
 We see that all these predicted spectra have a similar shape,
peaking at an energy of  $2-5$~MeV. A significant fraction of the neutrinos
in this peak can be attributed
to supernovae occurring at  $z\gtorder 1$,  whose emitted neutrinos are
subsequently degraded in energy due to the expansion of the universe.
The total integrated fluxes  corresponding to these calculations are;  
$5.4\ {\rm cm}^{-2}{\rm s}^{-1}$ (long-dashed),
$3.9\ {\rm cm}^{-2}{\rm s}^{-1}$ (high dotted),
$2.6\ {\rm cm}^{-2}{\rm s}^{-1}$ (short dash),
$2.2\ {\rm cm}^{-2}{\rm s}^{-1}$ (low dotted) and
$2.0\ {\rm cm}^{-2}{\rm s}^{-1}$ (solid).

It is instructive to compare our results with
 estimates of the neutrino spectrum which do not involve
the use of the $\Omega_g(z)$ evolution.
 The dot-dashed curve of Fig.3 shows the 
neutrino spectrum  we determine after approximating 
the supernova rate predictions of 
Totani {\it et al.} \cite{tot}, and is typical of the neutrino 
spectra predicted by them for a wide range of cosmological models.
For the time being we simply note that the
peak in our neutrino spectra is
 reduced by over an order of magnitude compared to the 
 calculations of \cite{tot}. Also, note that
the total integrated flux  corresponding this higher curve is
$34\ {\rm cm}^{-2}{\rm s}^{-1}$, typically an order of magnitude larger than 
the fluxes derived using $\Omega_g(z)$.

 In using the $\Omega_g(z)$ evolution to determine the neutrino spectrum, the
calculated flux 
is proportional to $H_{\rm 0}$.
The curves are plotted assuming $h=0.5$. Therefore, given the present
range of uncertainty in
  $H_{\rm 0}$, the
flux levels could be twice that shown. 
 There is also a dependence on the adopted
value of $\Omega_{\rm 0}$, through Eq.~(3). The spectra plotted are determined 
assuming $\Omega_{\rm 0}=1$. For $\Omega_{\rm 0}=0$, one finds the flux levels are
reduced by approximately a factor of $3$. However,  lower 
$\Omega_{\rm 0}$ models produce fewer metals for a given IMF \cite{me}. Therefore,
in such a model 
one can adopt a flatter IMF  in order to
normalize the metallicity at some redshift. This in turn
would lead to an increase in the predicted neutrino flux,  essentially
compensating
the reduction caused by altering the cosmological model.
This point illustrates the interplay between cosmological models
 and the cosmic IMF
-- one needs information on one in order to probe the other.

 In Fig.~4 the calculations are repeated for
the $\Omega_g(z)$ curve parameterizing the data of 
Lanzetta {\it et al.} \cite{LWT},
with the line-type of the curves retaining the meaning
ascribed to them for Fig.3. Here the $2-5$~MeV peak is more pronounced
as a consequence of the larger supernova rate at high redshift
predicted by this  $\Omega_g(z)$ evolution (essentially all the neutrinos in this
peak were produced at high $z$).
Again, relative to previous estimates, we see a deficiency in
the predicted relic neutrino spectrum at low energies.

For the dust-corrected evolution (dashed curve of Fig.1) the predicted 
neutrino spectra are very similar to those shown in Fig.~3. This is to be
expected since the predicted supernova rate derived using this $\Omega_g(z)$ 
evolution is similar to that predicted  by our standard model (see Fig.2).
 In general, we find from numerical experiments that 
simple models of 
dust  inclusion can affect the predicted flux levels by 
 factors of  up to 2. However, it is not clear at the present time
how large  the dust correction 
to the observed cosmic gas evolution should be
(see also \cite{loeb} for a discussion of 
gravitational lensing effects and dust). Thus,
we will say little more
about the  influence of dust.
The controversy over whether
or not
 it plays an important role in studies of damped Ly$\alpha$ systems
can be followed more in \cite{pett}.

Unfortunately, due to various  neutrino backgrounds in the MeV range
 (terrestrial, solar, atmospheric), the
relic supernova neutrino background is detectable by Superkamiokande
 only in the limited range
of approximately $15-40$ MeV. In this range the neutrino flux is 
largely dominated by the most recent supernovae, and the predicted 
 count rates in Superkamiokande are similar to those
reported in \cite{tot} (typically a few events per year).
Even if new detectors can go to lower thresholds, the large terrestrial
background of antineutrinos from nuclear reactors seems to
be an insurmountable problem \cite{Lagage}. This is a great pity since the most
interesting part of the relic neutrino background (i.e. that directly
connected to events in the early universe) will remain undetectable
by earth-based neutrino detectors. In order to explore the
low-energy supernova neutrinos some new indirect method needs to be developed.

Resonant absorption by ultra high-energy neutrinos on the relic 
supernova neutrino background is one such indirect method. 
This effect has previously
been discussed 
 in the context of
the relic cosmic neutrinos predicted by big bang cosmology
\cite{weil,Yoshida}.
However, although absorption processes on the relic supernova neutrinos
 can occur at   reasonable cosmic-ray energies, the low number abundance
of the  relic  MeV neutrinos removes any  reasonable
hope for this technique unless the peaks in the neutrino spectra of Figs.~3 and 4
were orders of magnitude larger.

Hopefully, the  importance of detecting relic supernova neutrinos in the
$2-5$~Mev range can inspire other, more hopeful, methods of detection.
 However, in the absence of such methods, other probes
of the supernova activity at high redshift deserve scrutiny.
It is possible that  cosmic-ray production at high redshift
could lead to spectral features in the diffuse gamma-ray background
 \cite{silk}.
By linking the past supernova rate with
past cosmic-ray activity, one could conceive of detecting such a
spectral feature 
commencing at $0.5 m_{\pi 0}(1+z_f)$ ($m_{\pi 0}=140$~MeV),
where $z_f$ marks the redshift at which significant galaxy formation begins.
 However, the gamma-ray production
from cosmic rays is maximized at $R_{SN}f(z)dt/dz$, where 
$f(z)$ depends on a myriad of uncertain effects such as
models of cosmic-ray propagation and confinement. 
This coupled with the
smooth rise  predicted for $R_{SN}$,  renders this an
unpredictable method for measuring the past supernova rate.
Below we  shall discuss alternative techniques 
of probing past supernova activity, which
appear more promising.

\medskip
\section{EVOLUTION  IN THE NUMBER OF EARLY-TYPE GALAXIES}

Let us further explore the discrepancy in the   low-energy
neutrino flux 
 predicted by our
 calculations compared to those reported by \cite{tot}.
We will argue  that  within the framework of \cite{tot},  
substantial evolution in the  number of early-type galaxies
would remove the discrepancy.

Totani {\it et al.} \cite{tot} determined a time-dependent
supernova rate derived from models of galaxy evolution based
on the population synthesis method. Deriving specific models of
galactic  evolution,
they calculated the supernova rate in
each  galaxy type.
Assuming a redshift at which galaxies form, 
and that the relative proportion of each galaxy
type (as measured at the present epoch)
remains constant with look-back time, a universal supernova rate was
determined. From this rate the
relic supernova neutrino background
was  determined in the same manner as outlined earlier.

Not surprisingly,
the higher neutrino fluxes at low energy predicted  by this alternative
 method
can be traced directly
to the supernova rates at early times
(cf. our Fig.~2 with Figs.3-5 of ref.~\cite{tot}). To further explore this,
let us adopt a particular cosmological model and compare
the  different supernova rates in detail.
We convert the  supernova rates
of \cite{tot} into redshift space, by
assuming a formation redshift of  $z_f=5$; which with
 $\Omega_{\rm 0}=1$ and $h=0.75$ leads to
 an evolutionary time of
$\sim 0.3$~Gyr at $z=3.5$. At this epoch the calculations of \cite{tot}
predict an average supernova rate of
 $\sim 1$ per year per $10^{11}{\rm M}_\odot$ galaxy.
(the predicted rate at this $z$
 would be higher for lower $z_f$ values.) 
Such high supernova rates are roughly $60$ and $20$ times
larger than the supernova
 rates predicted by the solid  and dotted curves, respectively,
 of Fig.2. 
Our calculations  show that for supernova rates 
this high,
the predicted gradient of the gas evolution ($d\Omega/dz$) at $z=3.5$
would be significantly steeper than that indicated by the data. This is shown better
 by considering the dot-dashed $\Omega_g(z)$  curve of Fig.1. 
The gradient of this curve at $z=3.5$ corresponds to  
a supernova rate
of  $1$ per year per $10^{11}{\rm M}_\odot$ galaxy, and is
clearly incompatible with the data.
Although for this calculation we adopted an IMF corresponding to $x=1.35$,
numerical experiments in which
we  vary the IMF within reasonable bounds do
not  alter the situation in any significant way.
It is evident  that
the relatively flat and smooth behavior of
 $\Omega_g(z)$ at high redshift severely limits the  allowed
supernova rate.

Galaxies which are elliptical (E) or intermediate 
between  elliptical and  spiral (S0),
totally dominate the supernova rate at high redshifts if their relative
abundance was equal to that at the present epoch.
 Therefore, if $d\Omega_g/dz$ remains relatively small
 one could conclude
 that the number of these galaxies must
decrease with increasing redshift. 
Exactly what reduction in the number of E/S0 galaxies
 would provide compatibility between the predicted and observed
 slopes of $\Omega_g(z)$ is difficult to
quantify. Clearly this would depend on  the true evolution
of $\Omega_g(z)$, and also on
 a rather uncertain error analyses of the presently
available data. However, 
for the 
cosmological model being discussed,  one can  say that the
 number of E/S0 galaxies  at $z\approx 3.5$
must be reduced by   over an order of magnitude
from their  abundance at the present epoch, in order to provide any sort
of match to the solid curve of Fig.1.
 Given the  closed-box assumption,
 this implies that the {\it number ratio}
 of E/S0 to other
galaxy types  must have decreased by this  same amount.

Note that $dz/dt$ decreases with decreasing redshift  -- this 
leads to two points.
First,
the gradient of the dot-dashed $\Omega_g(z)$ evolution (Fig.1)
  would have to be even steeper
at lower redshift in order to accommodate the  supernova rate of
$1$ per year per $10^{11}{\rm M}_\odot$ galaxy. Second,
a  redshift range $\delta z$ extends over longer evolutionary timescales at
lower redshifts. These effects offer the hope that
further sampling of the $\Omega_g(z)$ distribution at  low redshift
 would eventually
detect the onset of significant E/S0 formation. This would manifest itself
as a breakdown in the smooth extrapolation assumed, and would most likely
  be found
 in the last data bin at $z\ltorder 1.5$.

In principal, 
 the rapid gas evolution compatible with
the predictions of \cite{tot} could be occurring
prior to the highest observed redshift bin, and therefore  remain 
undetected. To accommodate this one must adopt much earlier values of
the formation redshift $z_f$, and/or a new
  cosmological model. Ideally one would prefer
  the first Gyr or so of
E/S0 galactic evolution to
occur at $z\gtorder 5$, such as would be the case for
   a low $\Omega_{\rm 0}$  model with $h=0.5$
 and $z_f\gtorder 7$.
If the relative number of E/S0 galaxies does not decrease
with redshift, there should be 
a very  rapid increase in $d\Omega_g(z)/dz$ for
$z\gtorder 5$. It remains to be seen whether this region will
 be   explored in the future.
 We note,  however, 
 there is tentative evidence
of a small
 {\it decrease} in the gas density at $z\gtorder 3$, possibly
 indicating  that above this redshift
 the largest damped Ly$\alpha$ systems have not fully formed
\cite{stor}.  If this decrease is confirmed,
it would further strengthen the case for the presence of
strong evolution in the E/S0 galaxy
population.

An increase  in the number of
 E/S0 galaxies with decreasing redshift
is predicted to occur in
 hierarchical galaxy formation models as a consequence of
merging processes.
In fact recently,  based on data from the CFRS survey \cite{lilly},
  evidence in support of such E/S0 evolution
has been put forward \cite{kauf}.
Contrary to this, however, is the argument 
 that the majority of the massive spheroid population
formed at $z\sim 3$. Several pieces of  observational evidence
support this view also, not least of which is the detection of large
elliptical galaxies at high redshift (see the review of
\cite{fuk}).
It remains to be seen whether future
survey statistics and numerical simulations
 will allow
 some combination of the two  differing E/S0 evolutionary scenarios
to be compatible with the evolution of the cosmic gas.

\medskip
\section{LATE INFALL MODELS}

Based on column density distributions, it has been argued
that the cosmic gas is best described by
a closed-box system \cite{LWT}. However, clearly it is difficult
to extrapolate such arguments within the last data bin of the observations.
We have already mentioned a possible discrepancy in our
predicted supernova rates at zero redshift (or more generally the
star formation rate) compared with what is inferred from observation.
We wish to point out a further discrepancy at low redshift between our
models and the observed data, which further suggests that
 closed-box models for the evolution of the cosmic gas 
require modification at late times.

In Fig.~5 we have plotted  as the solid curve
 our estimated rate of new metal production, $\dot\rho_z$,
assuming our standard model.  The data points are determined from
estimates of the global radiation flux below $3000$\AA , which is
assumed to be dominated by output from massive stars.
Details of the
UV emissivity to $\dot\rho_z$  conversion
procedures 
are discussed in \cite{madau}. The solid circles are based on Hubble 
Deep Field data \cite{madau},
whereas the open square is from ground-based data\cite{steidel}. It is important
to note that these represent lower limits to the inferred $\dot\rho_z$,
since it is plausible  that fainter and dust-obscured galaxies,
not included in these estimates,
are abundant at high redshift. The low redshift data (open circles) are
based on the CFRS data of \cite{lilly2}, and the current
value of $\dot\rho_z$  is based on the H$\alpha$ luminosity density of
the local universe \cite{Gallego}.

The predicted $\dot\rho_z$ from our standard model shows a consistent fit with
the high redshift data. Indeed, taken at face value it implies that the
Hubble Deep Field data is missing little from its inventory.
The  $\dot\rho_z$ values  predicted using the dotted curve of
Fig.~1 are factors of 2 and 6 higher at $z=4$ and $z=2.75$, respectively.
If the data points were to be  revised upwards  beyond these
levels,  a possible problem with our modeling
of $\Omega_g(z)$ at high redshift would be indicated.
At low redshift   a discrepancy between the
predicted and observed values of  $\dot\rho_z$ is clearly evident. Since 
$\dot\rho_z$ can be written as
\begin{equation}
\dot\rho_z=\rho_c 
  \int^{m_{up}}_{m(z)} m \Psi(z^f_m) {\Phi(m)} p(z^f_m)dm  \ \ \ ,
 \end{equation}
$p$ being the mass of the newly processed metals in the ejecta
divide by the initial stellar mass, then the discrepancy may once again
be related to a low rate of star formation at late times.
There are many ways to accommodate  this discrepancy, but perhaps
 the simplest is to allow for some infall of cool gas at late times.
 Dust-corrected models of the cosmic gas evolution which include   infall
that is a constant multiple of the instantaneous SFR,
 have been shown to be useful in this
regard \cite{fall}.
 The long-dashed curve of Fig.~5
shows the  $\dot\rho_z$ predicted from our  late-infall model. Here we have simply  added to
our evolutionary equations an infall term $I(t)$ where
\begin{equation}
I(t)=\Omega_g(z)\exp[-(|z-z_p|)^{n}]   \ \ \ {\rm Gyr}^{-1} \ \ \ ,
 \end{equation}
 where the fitting parameters $z_p$ and $n$ simply determine the
peak of the infall and its width in redshift space, respectively.
 The prefactor
$\Omega_g(z)$  assures that the infall is at  significant levels
at the chosen value of $z_p$. 
For the  dashed curve shown in Fig.~5 we have taken $z_p=1$ and $n=2$, and kept
the IMF of our standard model ($x=1.35$).
This  particular late-infall model is the one referred to in
 our earlier calculations of the supernova rate and the neutrino
background (Figs.2-4).
As is evident from  a cursory look at the evolution equations the addition of
any infall term would, for a given value of $d\Omega_g(z)/dz$, demand
a higher level of star formation. As can be seen from
  Fig.~5, the additional star formation associated with
the late infall clearly remedies the problem at low redshift.  Also, since the infall is at
 insignificant levels at higher redshifts, the closed box
results are reproduced at  $z\gtorder 2$. 

In Fig.~6 the long-dashed curve displays the
 effect the infall term of Eq.(8) has on the key metallicity tracer, Zn.
 For this calculation we assumed the infalling gas had a metallicity of $0.1$ solar
 (although even at this gas metallicity  the increased Zn yield   is almost completely determined by
the increased  SFR).
The solid curve corresponds to our standard model and the dotted curve
corresponds to the dotted curve of Fig.1. The data is from the tabulations of \cite{keck}
 (open circles) and \cite{pett2} (solid circle). It is clear that our infall term
does not spoil the consistency with the metallicity at $z\sim 2.5$. Although there
appears a discrepancy at  $z\sim 1$, we do not pay much credence to it because
the data
point here represents only three damped Ly$\alpha$ systems, and is most likely an
unreliable tracer of the global weighted average. Nonetheless, the prediction
of significantly higher
Zn abundances than those  indicated by the presently available  data 
at $z\sim 1$, is  a test of
our late-infall model.

\medskip
\section{DISCUSSION}

 It  appears then, that a model of the cosmic gas  described
by an essentially closed-box system at high redshift, but possessing significant
infall
 at lower redshifts,  can accommodate the available observational data.
 We stress that our infall term
 has not been modeled on any specific
large-scale structure simulation, but merely describes
the general form required in order to satisfy the observational constraints.
It is not at all obvious 
how the  complicated processes which could occur
 at $z\sim1$
 would lead an infall resembling that described by Eq.(8). 
This is material for a future study, but in the meantime let us make some 
remarks which may be helpful.

In studies of galactic chemical evolution, gas infall has been
shown to play an important role in accounting for the chemical and
photometric properties of spiral galaxies, eg. \cite{matt,arim}. 
It therefore seems plausible that infall onto galaxies is a universal phenomenon.
However,
in the context of the global cosmic gas, it is difficult to quantify 
the accretion rate of cool star-forming gas onto the damped Ly$\alpha$ systems. But
to satisfy our demands, it appears that this accretion must become
efficient only at $z\ltorder 2$. It is important to emphasize that with regard to  alterations
in the SFR
it is only 
the {\it net} infall that is important. It is likely that both inflow
and outflow processes are taking place simultaneously, and that the  net infall
 is the result of a combination of both
processes.

Based on our earlier results we suggested
 the number abundance of E/S0 galaxies
increases with time. The continuing merging processes associated with 
this evolution
 will lead to
alterations in the cosmic gas.
One would anticipate that the heat generated during mergers, eg. \cite{white},
 would lead to
an {\it outflow}  of hot  metal-enriched gas from the cool-gas
 system.
 Indeed, there is strong evidence from X-ray observations supporting the existence of a hot
metal-enriched intergalactic medium  that is correlated
with E/S0 galaxies\cite{mulch}. In addition, 
 if this hot gas represents a substantial fraction of the baryons
(there is every indication that it is at least comparable to the baryon fraction
 in stars), then ultimately
its existence must be accounted for in any model of the cosmic gas evolution. In our scenario
the hot gas 
would  have been
 produced recently due to the large E/S0 galaxy production occurring at low redshift.
The {\it net} outflow from  the cool-gas  system associated with  the hot gas production
could account for the rapid fall off in  $I(t)$ at $z<z_p$.
Beyond this  speculation,   one can hope that full blown  simulations, which include
all the necessary  input physics, will enlighten us further.

Finally, it is interesting
to note the connection between the chemical evolution of the cosmic gas
as studied here, and the models of Fields {\it et al.} \cite{field}.
This latter work,  which
outlines  some of the complexities involved
in modeling the merging processes of protogalactic gas clouds,
 attempts to account for metal-enriched intercluster gas and
the  most recent microlensing experiments \cite{alc}.
 A large remnant
white-dwarf population in the halo of our own galaxy seems to be the most rational
explanation
of the microlensing experiments.  If the same population was present
in
all spiral galaxies,
then effects on (or constraints from)
the cosmic gas may be anticipated. For example,
 our study shows how the gas should evolve under the influence of the very high
star formation rates that presumably accompany the formation
 of the white dwarf population.
 However, if a large fraction of
white dwarf formation took place
at very high redshift ($\sim 10$), as other observational constraints seem to require,
 then
compatibility with the observed evolution of the cosmic gas may 
still be possible. The  remnant white-dwarf  population would play no
role in the  subsequent $\Omega_g(z)$ evolution, and
any hot metal-enriched gas produced during the  star formation 
process
 would  play no role unless it cooled
and re-entered the star-forming gas. Nonetheless, some contamination from the white-dwarf
formation would have to exist at some level, and
 one could envisage a series
of metallicity indicators at redshifts of $z\sim3$
 (eg. plateau like behavior). 
Variation in the IMF at early times -- often invoked in discussions of the
microlensing results -- can also be probed by our calculations.
Although we always assumed a constant IMF, Fig.~5 indicates that IMF's
at early times with
significantly fewer high-mass stars 
 would not be consistent with  the metal
production at $z\sim 4$.
These too  are all topics
 worthy of further study, and  they highlight once again the importance 
of a detailed theoretical description of the  $\Omega_g(z)$ evolution.

\medskip
\section{CONCLUSIONS}

Based on the evolutionary history of the cosmic gas we have calculated the
past core-collapse supernova rate and the relic supernova neutrino background
that it gives rise to.
 We find  our predicted neutrino flux to
be substantially lower than other recent predictions -- a discrepancy which is
 largest at   lower energies.
 However, in the energy
 range accessible to 
Superkamiokande, $\sim 20$~MeV, the neutrino flux is found 
to be compatible with
previous estimates. This latter fact is not surprising since
the rate anticipated at Superkamiokande is largely influenced by
the supernova rate at $z\ltorder 1$. Novel techniques to
measure the relic background neutrinos at $E\sim 5$~MeV would be
of great importance as they would
provide a direct probe of star formation activity at $z\gtorder 1$.

Our predicted supernova rates have been used to
 provide some insight into the evolution of galaxies. 
 The discrepancy between
our predicted   rates  and previously reported rates, is removed if strong
evolution in the population of early-type galaxies is included.
Such an evolutionary trend seems  more consistent
with the evolution of the cosmic gas density, which appears to
constrain the birth of most E/S0 galaxies to either very early  ($z\gtorder 5$) or very late 
($z\ltorder 1.5$) epochs. We feel the latter epoch of formation to be more likely.

Although, a closed-box description for the evolution of the cosmic gas
seems to be  required by  observations at high redshift, we
suggest that an infall contribution at low redshift may be required.
This late infall term could accommodate  estimates of
the  global luminosity density evolution measured by recent  surveys.
It remains to be seen whether detailed numerical simulations and future observation
 will lead to the required form of
the late infall.

Finally, we have discussed possible connections between the 
evolution of the cool star-forming cosmic gas,
 and
 the creation of the hot metal-enriched
intergalactic gas detected by X-ray telescopes. 

 Investigations into the
chemical evolution of damped Ly$\alpha$ systems 
 offer new insights into the evolution of structure in the early universe.
With an ever increasing wealth of data on these systems being gathered,
  rapid progress
seems assured.

\medskip
\section{ACKNOWLEDGEMENTS}

I would like to thank  B. Chaboyer for input related to
 stellar yields, M. Fall and M. West for discussions, and the
anonymous referee for his comments.

\newpage

\section*{Figure Captions}

\bigskip
\noindent{\bf Figure 1:}
 $h\Omega_g$ as a function of redshift
for a de-acceleration  parameter $q_{\rm 0}=0.5$.
The solid circles
are data from\cite{stor}, and the open circles are from \cite{LWT}.
The curves are various parameterizations of the data,
 except the incomplete curve which is used to model very high
supernova rates (see text).

\medskip
\noindent{\bf Figure 2:} 
The supernova rate as a function of redshift. The different curves
correspond to the evolutionary curves of figure 1 (i.e. line markings
are same), except the long-dashed curve corresponds to a  model with infall.
An IMF slope
parameter of $x=1.35$ was used for the solid and long-dashed curves,
 with $x=1.7$ utilized
for the two other curves. The rates shown should be
 multiplied by $0.01/\Omega_{gal}$.

\medskip
\noindent{\bf Figure 3:} 
The $\bar\nu_e$ spectra at the present epoch,
calculated using solid $\Omega_g(z)$ evolution of Fig.1, with $x=1.35$
(except for dot-dashed curve). The different curves
correspond to different corrections  to the 
 supernova rate  at low redshift (see text). 
$\Omega_{\rm 0}=1$ and $h=0.5$ are assumed.

\medskip
\noindent{\bf Figure 4:} 
The $\bar\nu_e$ spectra at the present epoch,
calculated using dotted $\Omega_g(z)$ evolution of Fig.1, with $x=1.7$
(except for dot-dashed curve). The different curves
correspond to different corrections  to the 
 supernova rate  at low redshift (see text). 
$\Omega_{\rm 0}=1$ and $h=0.5$ are assumed.

\medskip
\noindent{\bf Figure 5:} The metal production density as a function
of redshift. The data points are  determined from 
\cite{madau} (closed circles), \cite{steidel} (open square), 
\cite{lilly2} (open circles),  and \cite{Gallego}  (closed square).
$\Omega_{\rm 0}=1$ and $h=0.5$ are assumed.

\medskip
\noindent{\bf Figure 6:} The Zn mass fraction in the cosmic gas
for our standard model (solid curve), infall model (long-dashed curve),
and model
assuming  the dotted  $\Omega_g(z)$ evolution of Fig.1  (dotted curve).
The data is from the tabulations given in \cite{keck}
 (open circles) and \cite{pett2} (solid circle).
$\Omega_{\rm 0}=1$ is assumed.

\vfill\eject
\tolerance = 1500
\baselineskip = 12pt
\font\bx =cmbx10 scaled \magstep2
\font\ti =cmti10 scaled \magstep2
\font\r  =cmr10  scaled \magstep2
\def\gsim{\raise.2ex\hbox{$>$}\kern-.75em \lower.7ex\hbox{$\sim$}}
\def\lsim{\raise.2ex\hbox{$<$}\kern-.75em \lower.7ex\hbox{$\sim$}}
\hoffset = .0cm
\hsize=6.5truein
\centerline{TABLE 1.}
\vskip .1cm
\vbox{
\centerline{Element Mass Fractions  (-log)}
\vskip.1truein
\hrule width 7.5truein
\vskip.03truein
\hrule width7.5truein
\vskip.1truein
 \halign{\hskip10pt#\hfil \tabskip.4truein  & #\hfil 
\tabskip.4truein & #\hfil \tabskip.4truein & #\hfil
\tabskip.4truein & #\hfil \tabskip.4truein & #\hfil \tabskip.4truein & #\hfil 
\tabskip.4truein & #\hfil \tabskip.4truein & #\hfil \tabskip.4truein & #\hfil
\tabskip.4truein & #\hfil
\tabskip.5truein & #\hfil\cr
\hfil z & \hfill Z & \hfil C & \hfil N  & \hfil O & \hfil Mg & \hfil Si &
 \hfil Ca & \hfil Cr & \hfil Fe & \hfil Zn \cr
        \openup1pt\cr
        \noalign{\hrule width 7.5truein}
        \openup3pt\cr

5.0 &  4.5 &  5.7 &  6.4 &  4.7 &  6.1 &  5.8 &  6.9 &  7.6 &  5.8 &  9.0 \cr   4.5 &  4.2 &  5.3 &  5.9 &  4.4 &  5.7 &  5.4 &  6.5 &  7.2 &  5.4 &  8.6 \cr 
4.0 &  3.8 &  5.0 &  5.5 &  4.0 &  5.4 &  5.0 &  6.2 &   6.9 &  5.1 &  8.3 \cr  3.5 &  3.5 &  4.6 &  5.1 &  3.7 &  5.0 &  4.7 &  5.9 &  6.5 &  4.7 &  8.0 \cr   3.0 &  3.2 &  4.3 &  4.7 &  3.4 &  4.7 &  4.4 &  5.5 &   6.2 &  4.4 &  7.7 \cr  2.5 &  2.8 &  3.9 &  4.3 &  3.0 &  4.4 &  4.0 &  5.2 &   5.8 &  4.0 &  7.3 \cr  2.0 &  2.5 &  3.6 &  4.0 &  2.7 &  4.1 &  3.7 &  4.9 &    5.5 &  3.7 &  7.0 \cr 1.5 &  2.2 &  3.3 &  3.7 &  2.4 &  3.8 &  3.4 &  4.6 &    5.2 &  3.3 &  6.6 \cr 
1.0 &  1.9 &  3.0 &  3.4 &  2.2 &  3.5 &  3.1 &  4.3 &   4.9 &  3.0 &  6.1 \cr  }

        \vskip.1truein
        \hrule width 7.5truein
        }
\vfill\eject

%\epsfbox{figure1.ps}
%\epsfbox{figure2.ps}
%\epsfbox{figure3.ps}
%\epsfbox{figure4.ps}
%\epsfbox{figure5.ps}
%\epsfbox{figure6.ps}

\end{document}